\documentclass[preprint,12pt]{elsarticle}

\usepackage{graphicx}
\usepackage{amssymb}

\usepackage{lineno} 

\journal{Physica A}

\begin{document}

\begin{frontmatter}


\title{A simple centrality index for scientific social recognition}



\author{Osame Kinouchi*, Leonardo D. H. Soares and George C. Cardoso}

\address{Departamento de F\'{i}sica, FFCLRP, Universidade de S\~ao Paulo\\ Ribeir\~ao Preto, 14040-901, Brazil}

\begin{abstract}
We introduce a new centrality index for bipartite network of
papers and authors that we call $K$-index. The $K$-index grows with the citation performance of the papers that cite a given researcher and can seen as a measure of scientific social recognition. Indeed, the $K$-index measures the number of hubs, defined in a self-consistent way in the bipartite network, that cites a given author. We show that the $K$-index can be computed by simple inspection of the Web of Science platform and presents several advantages over other centrality indexes, in particular Hirsch $h$-index. The $K$-index is robust to self-citations, is not limited by the total number of papers published by a researcher as occurs for the $h$-index and can distinguish in a consistent way researchers that  have the same $h$-index but very different scientific social recognition. The $K$-index easily detects a known case of a researcher with inflated number of papers, citations and $h$-index due to scientific misconduct. Finally, we show that, in a sample of twenty-eight physics Nobel laureates and twenty-eight highly cited non-Nobel-laureate physicists, the $K$-index correlates better to the achievement of the prize than the number of papers, citations, citations per paper, citing articles or the $h$-index. Clustering researchers in a $K$ versus $h$ plot reveals interesting outliers that suggest that these two indexes can present complementary independent information.
\end{abstract}

\begin{keyword}
Scientometrics\sep Hirsch index\sep Lobby index\sep complex networks\sep node centrality\sep citation network\sep Web of Science\sep social recognition\sep scientific prizes.\\
$^*$ Corresponding author.
\end{keyword}

\end{frontmatter}


\newpage

\section*{Highlights}
\begin{itemize}
\item We discuss centrality indexes in the bipartite scientific papers-authors networks;
\item We propose a new index (the $K$-index), which is simple to calculate and outperforms the Hirsch
$h$-index in several aspects;
\item The $K$-index correlates better with Nobel prizes 
than any of the other
indexes furnished by the \emph{Web of Science} (\emph{WoS}) platform;
\item The $K$-index is related to connections to the hubs of the scientific 
citation and social networks;
\item The outliers in the $K$ vs $h$ plane contain interesting 
information that can be used for scientific prizes prediction.
\end{itemize}

\section{Introduction}
\label{S:1}

Complex networks is an interdisciplinary area where statistical physicists have done impressive contribution~\cite{newmannetworks,barabasi2016network}. Due to an interest in complex networks with large and reliable data banks, physicists have frequently worked in Scientometrics, 
say in citation and scientific collaboration networks~\cite{newman2001structure,barabasi2002evolution,newman2004coauthorship,wang2008measuring,ren2012modeling,clough2016dimension,xie2016geometric}. Actually, one of the founding fathers of Scientometrics is the physicist Derek J. de Solla Price. A contribution that increased very much the interest of physicists on Scientometrics was the introduction of the $h$-index, also due to a physicist, Jorge E. Hirsch (Hirsch, 2005). Research in centrality indexes for citation networks experienced a boom after 2005~\cite{batista2006possible,egghe2006improvement,hirsch2007does,
schreiber2010twenty,todeschini2016handbook}. 

One of the decisive advantages of the $h$-index over its competitors is the ease of calculation even before platforms such as Web of Science (\emph{WoS}) and Google Scholar implemented it as an automatic feature. The centrality $h$-index has been used not only to evaluate individual researchers but also single papers~\cite{schubert2008using,martinez2014h}, 
scientific journals~\cite{braun2006hirsch,malesios2016measuring}, universities~\cite{abramo2013suitability} and countries~\cite{csajbok2007hirsch}.

	Despite its popularity, the $h$-index has some well-known 
    drawbacks~\cite{de2013h,bornmann2008there,waltman2012inconsistency,
    todeschini2016handbook}. An important one is that high impact scientists with a relatively small number of papers $N$ cannot have a high $h$ since $h \leq N$. Another issue is that scientists who clearly have different scientific recognition might have very similar $h$-indexes. One of the reasons for this ambiguity is that the $h$-index has a narrow dynamic range: rankings use integers from $0$ to about 
    $120$~\cite{hirsch2005index,garcia2012extension}. 
    
	The $h$-index is also prone to distortions due to 
    self-citations~\cite{de2013h,huang2011probing,schreiber2007self}. There are also issues related to dependence on the scientific area, although this can solved by proper normalization~\cite{batista2006possible,iglesias2007scaling,alonso2009h}. Lastly, it is not clear how the $h$-index correlates with more qualitative indicators of impact such as scientific 
    prizes~\cite{mazloumian2011citation,abramo2014authors}.
    
	Here we introduce a new index, which we call the $K$-index, designed to address the issues discussed above. The $K$-index does not depend on $N$, has large dynamic range and is very robust to self-citations. In addition, the $K$-index correlates better to scientific prizes than the number of papers $N$, the number of citations $C$, the average number of citations per paper $C/N$, the number of citing articles $CA$ and the $h$-index, that are the indexes readily available on \emph{WoS}. 
    
	Furthermore, in contrast to the vast majority of the bibliometric indexes discussed in the literature~\cite{todeschini2016handbook}, the $K$-index can be easily determined by simple inspection of the \emph{WoS} platform. Its calculation could potentially be automated on the \emph{WoS} in the future, in a manner similar to the calculation of the $h$-index.
    
	The $K$-index is a new node centrality index, related to but different from, the Lobby~\cite{korn2009lobby,campiteli2013lobby} and the $h(1)$ centrality indexes~\cite{lu2016h,pastor2017topological}. In addition, the $K$-index is competitive with the $h$-index, outperforming it in several important aspects.
    
\section{Material and Methods}
\label{S:2}
\subsection{The bipartite authors--papers complex network}
\label{SS:2.1}

First, we define the operator $\Theta$ that, when applied to
a weighted network $\bar{W}_{ij}$, generates a binary network 
$W_{ij} \in [0,1]$ following the Heaviside function 
$\Theta(\bar{W}_{ij}) =  W_{ij}$. 

Scientometric networks are not simple, but composed of several different layers and intermingled networks.
 For example, by only considering the open access data furnished by the \emph{WoS} platform, we obtain several complex networks:
 \begin{itemize}
 \item The author collaboration network $\bar{A}_{ij}$: this is
 an undirected weighted network where nodes (authors) $a_i$ and $a_j$
 are linked by the (integer) number $\bar{A}_{ij}$ of co-authored papers;
 \item The social collaboration network $A_{ij}$: the adjacency binary network $A_{ij} = \Theta(\bar{A}_{ij})$;
 \item The publication  network $P_{ik}$: the undirected binary matrix
 $P_{ik} = 1$ if author node $a_i$ published  paper node $p_k$ and zero otherwise;
 \item  The citation network $\bar{C}_{kl}$: the directed weighted 
 network where node paper $p_l$ cites (an integer number of times)
 $\bar{C}_{kl}$ the same paper $p_k$; 
 \item  The binary citation network $C_{kl}$: the directed binary network $C_{kl} = \Theta(\bar{C}_{kl})$;
 \item The self-citation network $S_{ikl}$: a directed binary matrix where 
 paper $p_l$ cites $S_{ikl}$ times the paper $p_k$ with the constraint that both are authored by $a_i$. This can be written as $S_{ikl}= 
 P_{ik} P_{il} C_{kl}$.  We observe that the self-citation network
 $S_{ikl}$ sometimes is blamed as a factor that inflates bibliometric indexes. This is a polemic topic,
 since a moderate level of self-citation can be a legitimate 
 scientific practice. 
 \item The citing articles network $C^A_{il}$:  the directed binary 
network $C^A_{il}$  where paper $p_l$ cites at least some paper authored by $a_i$. Notice that if paper $p_l$ cites two or more papers published by
$a_i$, this counts as a single citing paper, that is, $C^A_{il}=1$. This network matrix can be written as
 $C^A_{il} =  \Theta\left(\sum_k P_{ik} C_{kl}\right)$.
 \end{itemize}

\subsection{Centrality indexes in the WoS}
\label{SS:2.2}

We will use the following notation for the centrality indexes provided by \emph{WoS} for the $i$-th researcher :
\begin{itemize}
\item Number of papers $N_i = \sum_k P_{ik}$;
\item Number of citations $C_i = \sum_{k,l}  P_{ik} C_{kl}$ ;
\item Number of  citations without self-citations
 $C_i' = C_i - \sum_{k,l} S_{ikl} =  \sum_{k,l} P_{ik}(1-P_{il})C_{kl} $;
\item Number of citations per paper $C_i/N_i = \sum_{k,l}  P_{ik} C_{kl} / \sum_k P_{ik} $;
\item Number of articles that cite the researcher (or citing articles) 
$CA_i = \sum_l C^A_{il}$;
\item Citing articles without  self-citations 
$CA'_i = \sum_l (1-P_{il}) C^A_{il}$;
\item Hirsch index $h$;
\item Hirsch index $h'$ without self-citations.
\end{itemize}

\subsection{Calculation of the $K$-index}
\label{SS:2.3}

If a researcher has $K$ citing papers, each one with at least $K$ citations, then their $K$-index is $K$. By construction, the $K$-index measures not the raw quantity but the quality and importance of the citing articles: only highly cited papers that have cited the researcher enter in the calculation.

In contrast to most scientometric centrality  indexes~\cite{todeschini2016handbook}, the $K$-index is easily calculated by simple inspection of the \emph{WoS} platform. We assume that other platforms like Google Scholar Citations or Harzing Publish-Perish could be  easily adapted to provide $K$ without difficulty. On the \emph{WoS}, the procedure to get the $K$ of a researcher is the following:  
\begin{itemize}
\item Search the papers of a given author;
\item Click on the link \emph{Create Citation Report};
\item Click on the \emph{Citing Articles} ($CA$) link (or Citing Articles without self-citations ($CA'$), if desired);
\item Have the list of citing articles ranked from the most cited (defined as $r = 1$) to the least cited paper (this is the default ranking  presented by \emph{WoS});
\item Compare the citations $c(r)$ each citing article received to its rank $r$. In a descending order we have $r < c(r)$ up to a number $K$ where $K = r \leq c(r)$ but $K+1 = r+1 > c(r)$. This defines the $K$-index. 
\end{itemize}
The procedure is very similar to the calculation of the $h$-index, but now we use information from the the second layer of citations, i.e., citations to the citing papers $CA$.

\subsection{Correlation with other scientometric indexes}
\label{SS:2.4}

To study the correlation between the $K$-index and indexes $N$, $C$, $C/N$, $CA$ and $h$, we have sampled $28$ recent Nobel-laureate physicists and $28$ highly cited (but not laureate) physicists, the latter sample randomly chosen from the Thompson-Reuters page Highly cited Physics researchers 2014 (http://highlycited.com). For each researcher, we obtained $N$, $C$, $C/N$, $CA$, $h$ and $K$ from 
\emph{WoS}. Physicists from the past such as A. Einstein, P. Dirac and E. Ising, and also 2016 Nobel prize winners, were used for some comparisons but not for the study of correlations between indexes.

While $N$ is the raw productivity of a researcher and $C$, $C/N$ and $CA$ are generally considered as measures of impact, the $h$-index is viewed as a combination of productivity and impact. We propose that the $K$-index is associated to social recognition by scientific hubs. The $K$-index reflects citations by highly cited authors which probably are the 
hubs of the social scientific network $A_{ij}$ (or at least the authors of the paper hubs of the citation network $C_{kl}$). 


\section{Results}
\label{S:2}

First, we examine the correlation between the new $K$-index and the standard indexes extracted from \emph{WoS} for our two groups of scientists. Second, we study the correlation between the ranking produced by several indexes and Nobel prizes received by the sampled researchers.

\subsection{Comparison to standard indexes}

\begin{figure}[h]
\centering\includegraphics[width=1.1\linewidth]{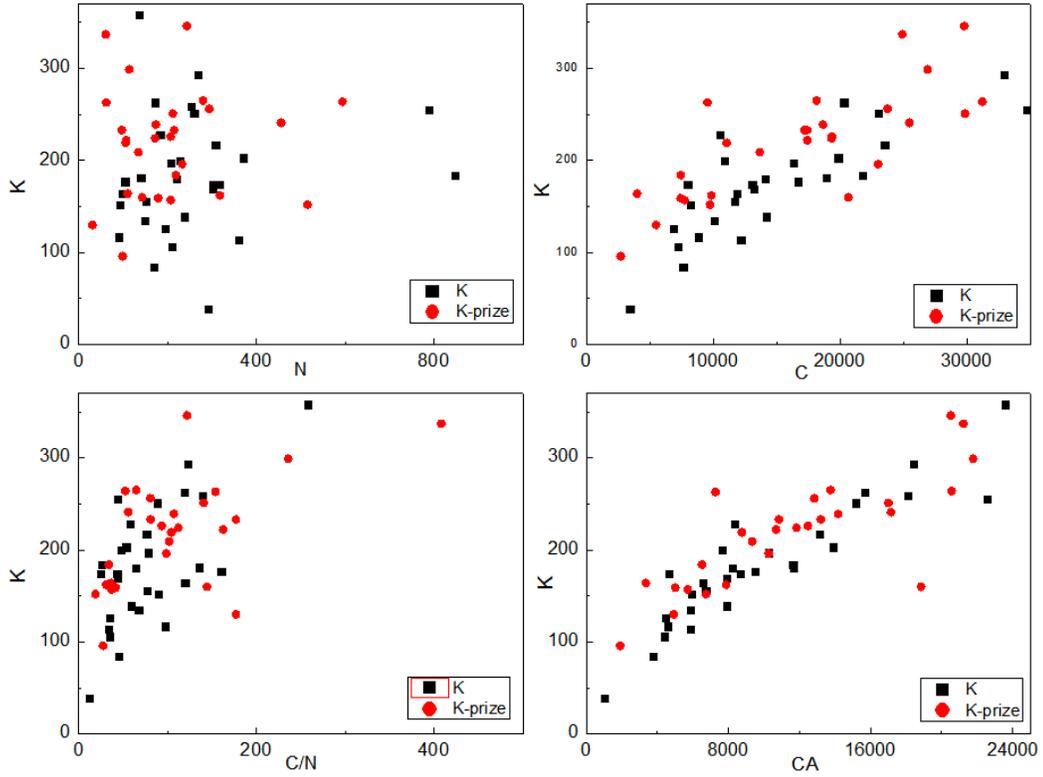}
\caption{{\bf Comparison between the $K$-index and standard \emph{WoS} bibliometric indexes:} a) $K$ versus $N$; b) $K$ versus $C$; c) $K$ versus $C/N$ and d) $K$ versus $CA$. The squares represent researchers who did not received Nobel prizes, while the circles represent laureates.}
\label{fig1}
\end{figure}

In Fig.~\ref{fig1}, we compare the $K$-index with the standard indexes furnished by \emph{WoS}. The correlations between $K$ and $N$ 
(Fig~\ref{fig1}a) and $K$ and $C/N$ are weak (Fig.~\ref{fig1}c), indicating that the $K$-index can give non-redundant information. There is a noticeable correlation between $K$ and $CA$ (Fig~\ref{fig1}d), but the same correlation happens for $h$ vs $CA$ (not shown), so this is not a particular weakness of the $K$ index.  Indeed, the correlation of Fig~\ref{fig1}d is somewhat misleading. For example, the $K$-index is much more robust to self-citations than $CA$. When self-citations are considerable, $CA'$s growth could follow an arithmetic progression –- of order $N^2$ –- while $K$-index depends only on the $K$-core of highly cited articles. Also, since $CA$ counts poorly cited, perhaps irrelevant, papers, it cannot detect fraudulent behavior, as we will discuss later in sub-section \ref{R5f}, where the researcher has high $CA = 1051$ but very low $K = 46$.

The correlation plots of Fig.~\ref{fig1} do not distinguish in a clear way between Nobel laureates and not laureates, that is, all researchers basically form a single cluster. In contrast, the $K$ vs $h$ plane brings unique information and suggests that laureates and non-laureates form different clusters, as we will discuss now.

\subsection{The $K$ versus $h$ plane}

\begin{figure}[h]
\centering\includegraphics[width=0.9\linewidth]{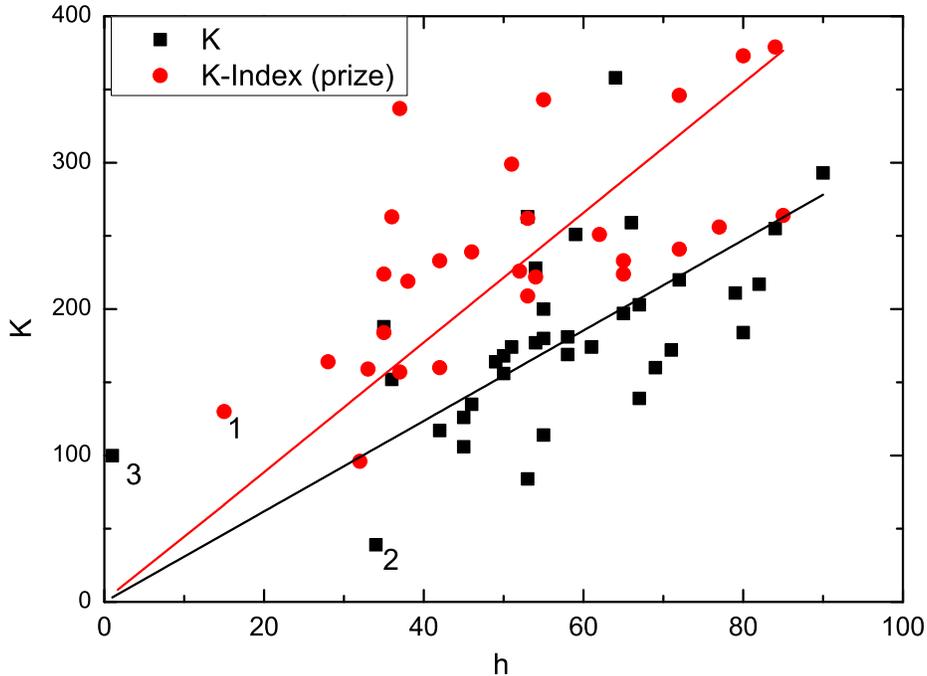}
\caption{{\bf $K$-index vs $h$-index for Nobelists (circles) and non-Nobelists (squares).} 1: Paul Dirac; 2: Mohamed El Naschie (scientist with controversial case of self citations); 3: Ernst Ising; The two lines are linear fitting curves for the Nobel Prizes and the non-laureate highly cited physicists that suggest the presence of two different clusters.}
\label{fig2}
\end{figure}

The $K$ and $h$ indexes seem to convey complementary information, as shown in Fig.~\ref{fig2}. Despite the overall correlation (Pearson correlation $= 0.7$), the interesting features are the outliers. We suggest the following classification, where the words true and false refer to the expectation created by the $h$ ranking:
\begin{itemize}
\item True positives (high $h$, high $K$): researcher with both high production, impact and recognition.
\item True negatives (low $h$, low $K$): researchers with low important production, impact and impact.
\item False negatives (low $h$, high $K$): researchers with few papers with high scientific recognition.
\item False positives (high $h$ low $K$): researchers with possibly inflated $h$ or with large production $N$ and high citation $C$, but with no exceptional scientific recognition.
\end{itemize}

As suggested by the different inclination of the curves in  Fig.~\ref{fig2}, for a similar $h$, Nobelists tend to have a higher $K$. However, there are four main types of anomalies: 
\begin{itemize}
\item Nobel laureates with low $K$ and $h$. This anomaly seems to be due to the area of research (say, a low average number of papers $N$)
and could be fixed by proper normalization~\cite{batista2006possible,iglesias2007scaling,alonso2009h}.
\item Non-laureates with very high $K$ and $h$. Here the anomaly could be either because Nobel prizes have some known biases (e.g., less prizes for pure mathematical physicists, despite high citations) or refer to non-laureated researchers with high prospects to be laureated in the near future.
\item False negative (in terms of $h$): the case of Ernst Ising (point $3$  in Fig.~\ref{fig2}, see subsection \ref{R5b}), who has $h = 1$ from a paper authored solo (Ising, 1925). Despite such a low $h$, $K$-index $= 100$ characterizing Ising as a pioneer on a fertile topic, achieving social
recognition by the hubs of the scientific social network.
\item False positive (in terms of $h$): the controversial case of El Nashie (point $2$ in Fig.~\ref{fig2}, see subsection~\ref{R5f}) illustrates the case of a false positive: $h = 35$ but $K = 46$ only.
\end{itemize}

\subsection{Nobel prize correlation curve}

We constructed a correlation curve between indexes and Physics Nobel prizes as follows: for each index ($N$, $C$, $CA$, $h$, $K$), we ranked the researchers of our sample. In the horizontal axis, we placed scientists from highest ($r = 1$) to lowest ($r = 56$) ranks. For each index, we build a ranked list of the researchers, from highest to lowest. Then, descending along the list, we compute the number $n(r)$ of Nobel Prizes found in researchers ranked up to rank $r$. The faster $n(r)$ grows the more sensitive and specific is the index for the identification of laureates. The curve $n(r)$ measures how many Nobel Prizes ($n$) occur up to the rank $r$ as defined by each index. 

\begin{figure}[h]
\centering\includegraphics[width=1\linewidth]{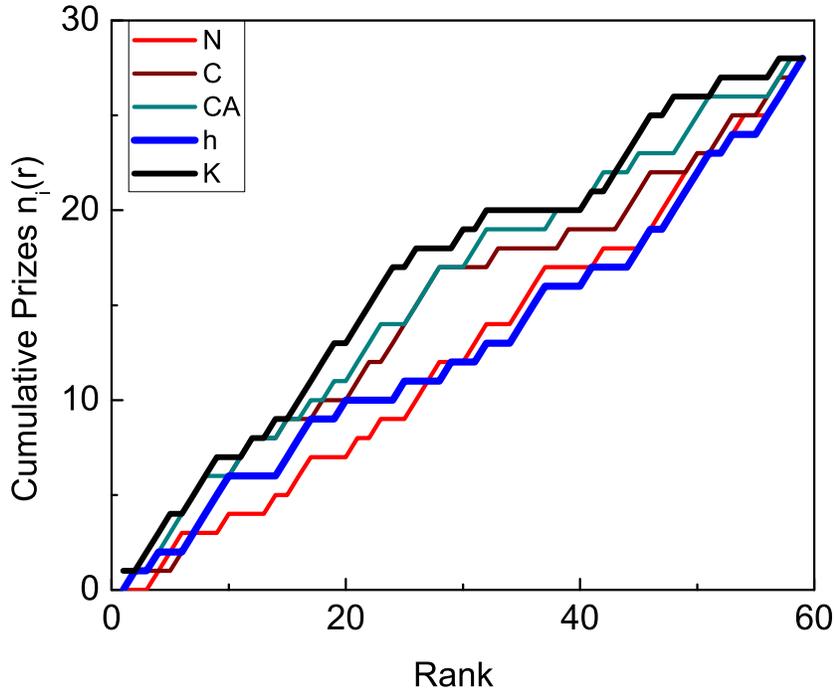}
\caption{ {\bf Cumulative curve $n(r)$.} Number of researchers $n(r)$ with Nobel prizes as a function of their relative rank $r$ for each index classification listed in the inset. The higher the value $n(r)$ for a given $r$, the better the correlation between the index and Nobel prizes.}
\label{fig3}
\end{figure}

The larger the area below the curve $n(r)$, the better the performance of the corresponding index. Fig.~\ref{fig3} contains one of the main proofs of superiority of $K$-index in scientific social recognition identification: if we rank researchers by the indexes $N$, $C$, $CA$, $h$ and $K$, we observe that the curve $n(r)$ for the $K$-index gives the best sensitivity and correlation with Nobel prizes. Interestingly, $K$-index performs well above the $h$ index. 

\section{Discussion: Why is the $K$-index superior to standard centrality indexes?}
\label{S:3}

\subsection{The $K$-index is very easy to compute}

An extensive and comprehensive compilation of bibliometric indexes can be found in a book by Todeschini and Baccini~\cite{todeschini2016handbook}. It is natural to ask if there is any gain in introducing yet another index such as the $K$-index. Indeed, the idea of using the second layer of citations has been advocated by Hu et al.~\cite{hu2011definition}, although not in the Hirsh-like form of the $K$-index. Indeed, one could argue that the $K$-index is related to the complex networks lobby index~\cite{korn2009lobby,campiteli2013lobby,rousseau2015general}. For this, we need to change the Lobby index original definition so that, instead of papers being single nodes, we collapse the total set of the researcher papers into a large and unique macro-node, and the citing articles would be considered as the other nodes in a (directed) network.

Our claim is not that there are no similar ideas in the literature. Our main claim is that the $K$-index, when compared to similar indexes, has an important and unique advantage: it is very easy to compute in the \emph{WoS} platform. Any researcher, by using \emph{WoS}, can calculate their $K$-index. The same cannot be said about most of the indexes, as can be seen in the aforementioned handbook~\cite{todeschini2016handbook}. We presume that, if the value of the $K$-index is recognized, \emph{WoS}, \emph{Google Citations} and other data banks can easily implement an automatic calculation of the $K$-index, as has already been done for the $h$-index.

\subsection{The $K$-index does not depend on $N$}
\label{R5b}

	We believe that one great advantages of the $K$-index is its indirect dependence on the number $N$ of papers published by the researcher. $K$ has no upper bound while $h \leq N$. An extreme example is Ernst Ising, who published an influential model~\cite{ising1925beitrag}, derived from his PhD thesis, in a single author paper. After that, due to Nazi persecution, Ising emigrated from Germany to Luxembourg and stopped to do any research. In 1947, he emigrated to the United States and becomes a professor at Bradley University. 
	His 1925 paper received $1373$ citations at the time of this writing. The only other paper Ising would publish, indeed a commentary at the \emph{American Journal of Physics} about Goethe's optics~\cite{ising1950goethe}, has a single citation in \emph{WoS}.  Therefore, Ising $h$-index is equal to one (by the way, with our present citation, Ising $h$-index will grow to its maximum value $h = 2$). 
	It seems unfair to give such a low ranking to someone who published the pioneering idea on what is now known as the Ising model. The $h$-index puts Ising in the same ranks of an undergraduate student with a single paper cited a single time. The $K$-index tells a very different story: Ising has a considerable value $K = 100$. This is fairer and puts Ising well above students with single papers.  The $K$-index makes justice to Ising, indicating his scientific social recognition. 
	Another example is Paul Dirac, who has a low $h = 15$. However, he has a $K = 130$, signaling also social recognition by other scientists (indeed, there are four Physics Medals in honor of Dirac). This low value for $h$ is not due to the fact that Dirac is a scientist from the beginning of the 20th century, but seems to be correlated to his lower productivity $N$, which is a constraint for his $h$-index.

\subsection{The $K$-index has larger classification range than $h$-index}

	As observed in the $K$ vs $h$ plane of Fig.~\ref{fig2}, the range of $K$ is at least a factor of three times larger than the range of $h$. This implies that $K$ produces a finer discrimination between researchers. This is a clear advantage for ranking purposes. On the other hand, the $K$-index continues to be based on integers. This feature gives more stability to ranking since real numbers ranks are more prone to spurious ranking order due to small statistical fluctuations.

\subsection{$K$-index better discriminates scientists with the same $h$-index: the Einstein-Hirsch paradox}

The $h$-index frequently presents paradoxical results on its ranking, giving similarly large $h$ to researchers with very different scientific recognitions. For an example, Jorge Hirsch's  $h$-index is $55$, while Albert Einstein's is $51$.  It seems that, due to its limited classification range, the $h$-index has poor discrimination power for researchers with large $N$ and $C$.
The $K$-index for Hirsch is $200$, which is high, but Einstein's is superior: $K = 299$.  This is another demonstration of how $K$-index can be a powerful tool to discriminate between scientists of very different scientific social recognition and similar $h$.
Let us take a look at another example:  Einstein ($h = 51; K = 299$) and Edward Witten, the most cited physicist of the world ($h =120; K = 368$). While Witten’s $h$ is $2.35$ times Einstein’s $h$, Witten’s $K$ is only $1.23$ times Einstein's $h$, what seems to be more fair in terms of scientific social recognition.

\subsection{The $K$-index is very robust to self-citations}

Several studies have shown that the $h$-index is prone to manipulation by self-citations~\cite{schreiber2007self,huang2011probing,viiu2016theoretical}. It is very easy to see why the $K$-index is robust to self-citations. Only highly cited papers compose the $K$-core.  If a researcher makes a self-citation, it will only be counted for $K$ if it is of an already highly cited paper.  Of course, some difference $\Delta K = K - K'$ may appear between the $K$-index calculated from all citing articles and the $K’$-index due to citing articles without self-citations (also furnished by \emph{WoS}). Our claim is that $\Delta K/K$ is much smaller than the corresponding $\Delta h/h$. \emph{En passant}, we notice that the $K'$-index ($K$ calculated without self-citations) is easily obtainable from \emph{WoS} while the $h'$-index ($h$ calculated without self-citations) is not furnished or directly 
calculable in the platform.

\subsection{The $K$-index detects scientific career frauds}
\label{R5f}

A public instance of scientific fraud has been denounced by 
\emph{Nature}~\cite{schiermeier2008self}. The researcher, El Naschie, who was the editor of a scientific journal, was suspected of scientific misconduct. Superficially, his scientific career looked impressive: $N = 293$ and $h = 35$, which would put him among top ranking physicists. We could compare El Nashie with known physicists such as Robert H. Swendsen ($N = 149; h = 37$) and Ronald Dickman ($N = 134; h = 28$). 
However, for El Naschie we have only $K = 46$, in contrast to $K = 206$ for Swendsen and $K = 76$ for Dickman. Also, the $K$-index calculated by using the $CA$ without self-citations is $K´ = 39$ (El Naschie), $K´ = 204$ (Swendsen) and $K´ = 71$ (Dickman), showing that, proportionally, the $K$ index is much more affected by self-citations in the abnormal case: $K/h = 1.31, \Delta = (K-K')/K' = 0.18$ (El Naschie), $ K/h = 5.57, \Delta = 0.01$ (Swendsen), $K/h = 2.71, \Delta = 0.07$ (Dickman). We suggest that unusually low $K/h$ and $K/N$ and a higher $\Delta$ ratios could be indicative of abnormal publication patterns. 

\subsection{The $K$-index correlates well with scientific prizes}

Some literature studies have discussed the relationship between centrality indexes, specifically the h-index, and scientific success in terms of future performance and recognition such as scientific prizes~\cite{hirsch2007does,vinkler2007eminence,honekopp2012future,penner2013predictability}.
The $K$-index correlates better with Nobel prizes than $N$, $C$, $C/N$, $CA$ and $h$, as shown in Fig.~\ref{fig3}. Indeed, it is surprising that the $h$-index performs worse for this task than any of the other indexes compared. In addition, the $K$-index has the smallest coefficient of variation (CV) for Nobelists, see Table~1. This means that the $K$-index correlation with Nobel Prizes is tighter than that of the other indexes.

\begin{table}[h]
\centering
\begin{tabular}{l l l l}
\hline
\textbf{} & \textbf{K (prize)} & \textbf{CA (prize)}& \textbf{h (prize)}\\
\hline
Mean & 224 & 12792 & 52\\
Standard deviation & 66 & 8286 & 18 \\
CV & 29\% & 65\% & 35\% \\
\hline
\end{tabular}
\caption{\bf{Statistics for the sample of laureates for different indexes.} The K-index has the lowest coefficient of variation (CV) among competing indexes.}
\end{table}

Here we must make a disclaimer: correlation is not prediction. We cannot to say that a high enough $K$-index predicts that the researcher will win a prize. Our test sample included researchers who received the prize in the last 18 years, most of them in the last 10 years; probably $C$ and $CA$ increase after a prize –- the pattern was not clear from the data. To study $K$ as a predictor of scientific prizes we must count only the citations received by a laureate scientist before the prize (or study scientists that have not yet been awarded a prize and follow up for possible prizes).
.An illustrative example is given by the 2016 Physics Nobel Prizes. By using their metrics, \emph{WoS} made a list of probable winners a few days before the announcement of the prize. It seems that \emph{WoS} metrics correlates well with the $h$-index, since all selected researchers have a very large $h$: J. A. York ($h = 79, K = 211$), R. K. S. Thorn ($h = 72, K = 220$), C. Grebogi ($h = 69, K = 160$) and W. Drever ($h = 50, K = 168$). However, the true winners have a lower $h$ with a higher $K$ (all above $K = 200$): D. J. Thouless ($h = 55, K = 343$), F. D. M. Haldane ($h = 53, K = 262$) and J. M. Kosterlitz ($h = 35, K = 224$). This anecdotal evidence suggests that ranking by using $K$ seems to better predict the Nobel Prize than ranking by $h$. Of course, this must be checked by using larger samples.  Our point is that for the 2016 Nobel Prize, the $K$-index has passed the test when compared with the \emph{WoS} prediction: the actual winners had lower $h$ than most of the researchers predicted by \emph{WoS}, while all had higher $K$ than those of the \emph{WoS} selection. 

\subsection{Limitations of the $K$-index}

Since $K$ does not depends on $N$, it is a scientific impact index not a productivity index. As we have seen, Ernst Ising with $N = 2$ has a large $K$-index of $100$. Someone with few papers that publishes (possibly with seniors researchers) a well-cited review or participates in large science collaborations might have a large $K$ and a small $h$. The outliers with small $h$ and high $K$ in the $K$ vs $h$ plane must be carefully analyzed. Some are big science low $N$ collaborators, some are authors of reviews, but others are low $N$ high impact scientist, even Nobel prize winners (Dirac). Inflation of $K$ due to publishing reviews can be mitigated by eliminating the reviews in the researcher's $WoS$  statistical summary before searching for  citing articles $CA$, but the other outlier instances must be examined case by case.

The $K$-index does not discriminate the following (extreme) situation. Consider two researchers, say Alice with $h = 30$ and $K = 80$, where her $K$ is due to citations for all her papers. The other researcher, say Bob, also has $h = 30$ and $K = 80$, but now his $K$ is due to citations to a single important paper. Although artificial and rare, this extreme case could occur and $K$ does not discriminate Alice from Bob. However, the $h$-index can not distinguish between them either. 

This particular example only confirms a trivial fact in Scientometrics: to fully characterize the bibliographic production of a researcher, instead of a small set of scalar indexes, we need the full networks $P_{ik}, \bar{C}_{kl}, A_{ij}$ of the bipartite network of scientific papers and authors. And, if we want to compare researchers from different scientific areas, some kind of normalization (in $N$, $C$, $C/N$, $CA$, $h$, $K$ etc.) must be done so that the distributions collapse into universal curves as shown in~\cite{batista2006possible,iglesias2007scaling,alonso2009h}. 

\subsection{Comparison with other new indexes}

In this paper we compare our centrality $K$-index only with the standard indexes  $N$, $C$, $C/N$, $CA$ and $h$. No comparison is made with new indexes such as the $g$ index~\cite{egghe2006improvement} and the $h_I$ individual Hirsch index~\cite{batista2006possible}. We chose to work with the \emph{WoS} database, where the $K$-index can be easily determined, while the computation of indexes such as $g$ index and  $h_I$ is not trivial. We defer a comparison between the $K$-index and such indexes to another paper.

It is important to note that, although inspired by, the $K$-index is not equivalent to the Lobby~\cite{korn2009lobby,campiteli2013lobby} or the $h(1)$~\cite{lu2016h,pastor2017topological} indexes applied to citation networks. The $h(1)$ index, as a natural extension of the $h$ index (called $h(0)$), would be determined as follows: a researcher has index $h(1)$ if they have $h(1)$ papers with $h(1)$ citations, each one with at least $h(1)$ citations. This is very different from the way $K$-index is defined. In addition, $h(1)$ cannot be readily determined by quick inspection in the \emph{WoS} platform as the $K$-index can. 

Out of $150$ bibliometric indexes listed in~\cite{todeschini2016handbook}, none of them can be easily calculated by inspection the \emph{WoS}, with the exception of the $h$-index.  This fact alone makes the $K$-index superior, in the sense of easy of determination, to all those centrality indexes. And easy calculation is a pivotal property for a centrality index to become accepted and popular.

\subsection{Generalizations of the $K$-index}

As has been done to other indexes, we can generalize and adapt the $K$-index to evaluate research impact in a number of contexts:
\begin{itemize}
\item {\bf Proximal $K_m$ index:}  in the calculation of the $K$-index, only articles from the researcher published in the last $m$ years are considered. This can be easily done in the \emph{WoS} by selecting the time interval of the search.
\item {\bf Recent impact $K_y$ index:} in the calculation of the $K$-index, only citing articles published in the last $y$ years are accounted. This can be done in the \emph{WoS} by sorting the citing articles by year, eliminating those published before the last $y$ years, then sorting anew by citation importance and calculating the $K$-index as before.
\item {\bf Group $K_g$ index:} As in the case of an individual researcher, we can compute the citing articles received by a group (research team, department, university, country etc.). For department, university and country it is currently possible for open access users of the \emph{WoS} to compute the $K$-index if the number of records is lower than $10,000$. We can work around this limitation if we select the search for a limited number of years for departments and universities, and for single years (or semesters) for countries. So, we can also obtain a $K_g(t)$ index that is a function of time.
\end{itemize}

\section{Conclusion and Perspectives}
\label{S:4}

We have introduced a new citation index (the $K$-index) that, like the $h$-index, is easily obtainable from the \emph{WoS} platform. The $K$-index outperforms the Hirsch index in several aspects: has a larger classification range, distinguishes researchers with similar $h$, does not depend on the number $N$ of papers, is robust to self-citations, is a good detector of cases of some scientific frauds and has better correlation to scientific prizes. 
The plane $K$ vs $h$ also gives interesting information, as exemplified by the data for the 2016 Nobel Prize. The study of this plane as a predictor of scientific prizes seems to be very promising, by using other samples such as the Wolf prize, the Boltzmann medal, and Maxwell medal.

Acknowledgments:
O. K. acknowledges financial support from the CNAIPS – Center for Natural and Artificial Information Processing Systems (USP). This paper results from research activity on the FAPESP Center for Neuromathematics (FAPESP grant 2013/07699-0). G. C. C. acknowledges funding from CAPES project number 88881.067978/2014-01.






%

\end{document}